\documentclass[11pt,a4paper]{article}

\usepackage[utf8]{inputenc}
\usepackage[T1]{fontenc}
\usepackage[english]{babel}

\usepackage[a4paper,margin=25mm]{geometry}

\usepackage{amsmath}
\usepackage{amssymb}
\usepackage{amsfonts}

\usepackage{graphicx}

\usepackage[authoryear,round,sort]{natbib}
\usepackage[hidelinks]{hyperref}

\title{\bfseries A One-Dimensional Integral Equation for a Porous Horizontal Disc under Water Waves}

\author{
  Luiz Fernando de Moraes Campos Filho\,$^{1}$ \quad
  Leandro Farina\,$^{2}$ \quad
  Juliana Sartori Ziebell\,$^{2}$ \\[2mm]
  \small $^{1}$Federal Institute of Mato Grosso, MT, Brazil \\
  \small $^{2}$Institute of Mathematics and Statistics,
  Federal University of Rio Grande do Sul, RS, Brazil \\[1mm]
  \footnotesize
  \href{mailto:campos.filho@ifmt.edu.br}{campos.filho@ifmt.edu.br} \quad
  \href{mailto:farina@mat.ufrgs.br}{farina@mat.ufrgs.br} \quad
  \href{mailto:julianaziebell@ufrgs.br}{julianaziebell@ufrgs.br} \\[1mm]
  \footnotesize ORCID:
  Campos Filho \href{https://orcid.org/0009-0006-4600-4914}{0009-0006-4600-4914};
  Farina \href{https://orcid.org/0000-0003-2744-515X}{0000-0003-2744-515X};
  Ziebell \href{https://orcid.org/0000-0001-8244-5051}{0000-0001-8244-5051}
}

\date{}

\begin{document}

\maketitle

\begin{abstract}
\noindent
Wave scattering by a thin, porous circular plate submerged in deep water is
investigated. The problem is formulated as a second-kind hypersingular Fredholm
integral equation over the unit disk, solved numerically using the Boundary
Element Method. The analysis focuses on calculating hydrodynamic forces,
specifically added mass (real part) and damping coefficient (imaginary part).
Results demonstrate the influence of the porosity parameter $G$: less porous
plates ($G$ real) increase added mass and hydrodynamic force, while more porous
plates ($G$ imaginary) reduce these effects but increase the damping
coefficient. The proposed formulation is validated, showing excellent agreement
with established literature.

\end{abstract}

\begin{center}
\begin{minipage}{0.92\textwidth}
\small
\noindent\textbf{Resumo.}
O espalhamento de ondas por uma placa circular fina e porosa submersa em águas
profundas é investigado. O problema é formulado como uma equação integral
hipersingular de Fredholm do segundo tipo sobre o disco unitário, resolvida
numericamente pelo Método de Elementos de Contorno. A análise se concentra no
cálculo das forças hidrodinâmicas, expressas em termos de massa adicional
(parte real) e coeficiente de amortecimento (parte imaginária). Os resultados
demonstram a influência do parâmetro de porosidade $G$: placas menos porosas
($G$ real) aumentam a massa adicional e a força hidrodinâmica, ao passo que
placas mais porosas ($G$ imaginário) reduzem esses efeitos e elevam o
coeficiente de amortecimento. A formulação proposta é validada por meio de
excelente concordância com a literatura.

\medskip
\noindent\textbf{Palavras-chave:} Disco Poroso; Massa Adicional; Ondas de Água;
Coeficiente de Amortecimento; Equação Hipersingular.
\end{minipage}
\end{center}

\bigskip

\section{INTRODUCTION}

The study of wave interaction with a submerged horizontal body has been the object of great attention from researchers in the field of marine hydrodynamics and coastal engineering. A submerged horizontal platform can serve as an essential component in various coastal and offshore structures. Its application ranges from breakwaters \citep{xiping94b} and wave energy converters (WECs) \citep{astariz15} to coastal barriers, very large floating structures (VLFS) \citep{Lamas}, and semi-submersible floating offshore wind turbines (FOWT) \citep{antonutti14, LopezSouto}. The motivation for its use stems from the versatility and effectiveness of this configuration in different maritime contexts.

According to \cite{farina12}, the relevance of these studies is justified, as the evaluation of the hydrodynamic force acting on the floating and/or submerged body is fundamental in many problems of interest to ocean engineers and naval architects. For example, in industrial, scientific, commercial, and military activities at sea, it is important to understand the influence that waves exert on large floating or submerged structures in the water.

A range of possible cases exists in the study of wave interaction with submerged bodies, considering the possibilities of the geometry and material of the body under analysis. In this work, studies related to cases where the bodies are thin rigid plates or thin porous plates are presented.

The study of wave interaction with submerged rigid bodies/plates has been carried out by various authors. A starting point is the problem for a floating obstacle, also known as the \textit{dock problem} \citep{garrett1971}. Such a problem can be reduced to solving a second-kind Fredholm boundary integral equation for the velocity potential. \cite{islam} studied the scattering and radiation of water waves by a submerged rigid disc in a two-layer fluid, reducing the problem to a one-dimensional second-kind Fredholm integral equation, and the results were presented in terms of the hydrodynamic force. \cite{das} conducted the same study, but in a three-layer fluid. Parsons and Martin, in a series of publications (\citealp{PARSONS1992}; \citealp{PARSONS1994}; \citealp{Parsons1995}; \citealp{martin1997interaction}) investigated several two-dimensional water wave problems, reducing the study to hypersingular integral equations. These investigations addressed scattering problems by submerged flat discs \citep{PARSONS1992}, curved plates and plates emerging at the surface \citep{PARSONS1994}, and water wave trapping by submerged plates \citep{Parsons1995}. They utilized an expansion-collocation method to solve the one-dimensional equations, employing Chebyshev polynomials of the second kind for the expansion in \cite{PARSONS1994}. \cite{ZIEE} studied the submersion of a thin, rough, circular disc in a deep-water free surface. The problem was reduced to a hypersingular equation over the body's boundary, and the results were presented in terms of the hydrodynamic force. A study similar to that of \cite{ZIEE} was presented by \cite{FARINA2017165}. A key methodological distinction lies in the plate geometry: \cite{ZIEE} investigated thin, rough plates, while \cite{FARINA2017165} focused on thin, smooth, and nearly circular plates. Additionally, the mathematical approaches for solving the problem also differ: \cite{ZIEE} employed the perturbation method to obtain the governing equation, whereas \cite{FARINA2017165} utilized conformal mapping. In both works, the results were presented in terms of the hydrodynamic force.

The interaction of waves with porous plates in two dimensions has also been the subject of investigation by various authors, and the results are presented in terms of wave reflection and transmission. \cite{Sollit} presented the first theoretical research on wave propagation in a porous medium, where reflection and transmission coefficients for a permeable breakwater of rectangular cross-section were predicted. The mathematical model is developed by expressing the normal fluid velocity through the porous plate as proportional to the jump in the velocity potential across the plate \citep[eq 2]{chwangat94}. This relationship is based on the assumption that the flow in the porous medium is governed by Darcy's law \citep{chwang1983}. Subsequently, singular integral equation methods have been applied to two-dimensional wave-porous plate problems; Gayen and Mondal used Chebyshev polynomial-based expansion-collocation methods to solve the governing hypersingular equations for a submerged porous plate \citep{gayen14} and for two symmetric inclined permeable plates \citep{gayen16}, while \cite{Koley} considered the scattering of waves by a floating flexible porous plate using Fredholm integral equations and obtained solutions using Simpson's quadrature formula.

According to \cite{farina21}, the investigation of the interaction of three-dimensional waves with porous plates has received less attention to date and has mainly relied on the eigenfunction expansion method. \cite{chwangat94} and \cite{Liu11} addressed the phenomenon of wave scattering by a submerged horizontal porous disc. The former study discussed the non-dimensional surface elevation and the vertical wave force. The latter study presented results related to the non-dimensional wave height concerning water depth and disc radius, in addition to the vertical force exerted on the disc. On the other hand, \cite{Molin} analyzed the axisymmetric problem of water wave propagation around a submerged porous disc, providing data on the added mass and damping coefficients as a function of the Keulegan-Carpenter number. Recently, several studies \citep{dokken17, Ouled, Mackay} have focused on the diffraction and radiation of waves by bodies with porous components, providing data on the added mass and damping coefficients, although they did not specifically address the case of porous plates.

The study of porous plates introduces a parameter of great importance, the {\bf porous effect parameter}, represented by $G$. The works by \cite{xiping95}, \cite{xiping94b}, \cite{YuC94} introduced a boundary condition for thin porous plates, in which $G$ depends on the resistance force ($f$), the inertial coefficient ($S$), and the geometric properties of the porous plate. Since $G$ is a complex number, it is defined as
\begin{equation}\label{f4}
\centering
G=\frac{\gamma(f-iS)}{K\bar{b}(f^{2}+S^{2})}=G_{r}+iG_{i},
\end{equation}
where $\gamma$ is the plate porosity, defined as the ratio between the volume of the porous part and the total volume, $\bar{b}$ is the physical thickness of the porous plate, and $K=\omega^{2}/g$, with $g$ being the acceleration of gravity and $\omega$ the wave angular frequency.

\cite{das22} investigated the radiation and scattering of flexural gravity waves by a submerged porous disc in a deep ocean with ice cover, assuming linear theory. The problem was reduced to the solution of a hypersingular boundary integral equation that can be further reduced to a system of one-dimensional Fredholm integral equations of the second kind. The hydrodynamic force for the scattering problem, and the added mass and damping coefficients for the radiation problem, were determined and numerically computed.

In this work, the wave radiation by a submerged porous disc in a free surface fluid, but without the ice layer, will also be investigated. Although the work presented by \cite{das22} is more general, this research employs a different method. Following the steps of \cite{farinamartin97}, the governing integral equation is reduced to a one-dimensional integral equation under axisymmetric motion. We employ the Boundary Element Method with robust numerical integrations to validate the integral formulation against results from \cite{farinamartin97} and \cite{farina21}. The added mass and damping coefficients will be analyzed, with new numerical results detailing their dependence on the wavenumber for different porosity coefficients $G$ and different depths $d$.

\section{METHODOLOGY}

\subsection{Formulation}

\begin{figure}[htb!]
\centering
\includegraphics[scale=0.55]{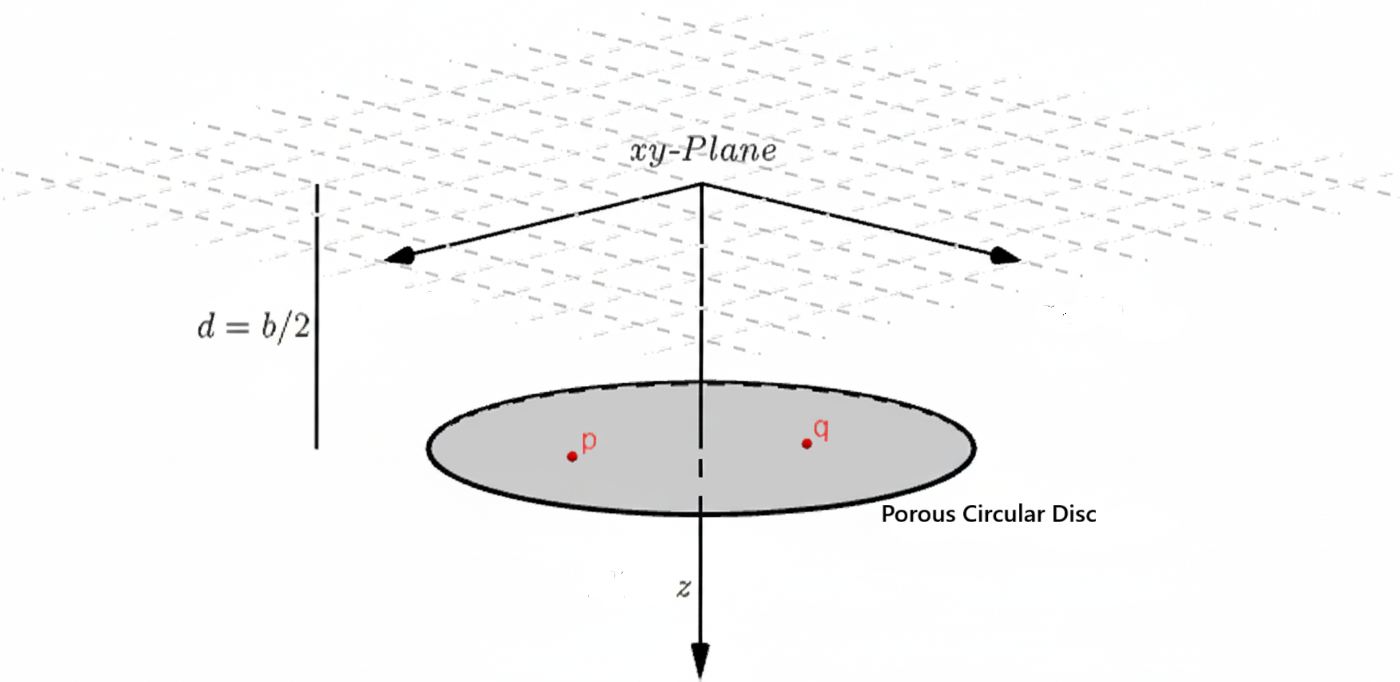}
\caption{Geometry of the problem studied.\label{im1}}
\end{figure}

Consider a Cartesian coordinate system $(x,y,z)$, where $z<0$ and $z=\zeta(x,y,t)$ defines the free surface elevation. Also consider that a body, with surface $D$, is fully submerged below the free surface of a given fluid; $D$ is a circular, porous and closed surface, as presented in Figure \ref{im1}.

Assume that the fluid motion is of small amplitude, irrotational, incompressible and inviscid to allow the introduction of a velocity potential $Re\left\{\phi(x,y,z)e^{-i\omega t}\right\}$, where $\phi$ satisfies the following conditions, as presented in \cite{farina12}
\begin{equation}\label{f1}
\left(\frac{\partial^{2}}{\partial x^{2}}+\frac{\partial^{2}}{\partial y^{2}}+\frac{\partial^{2}}{\partial z^{2}}\right)\phi_{j}=0,
\end{equation}
\begin{equation}\label{f2}
\frac{\partial \phi_{j}}{\partial z}-K\phi_{j}=0,~j=1,\ldots,7,
\end{equation}
\begin{equation}\label{f3}
\frac{\partial \phi_{j}}{\partial n}=V_{j}+iKG\left[\phi_{j}\right],~on~D,
\end{equation}
where $K=\omega^{2}/g$, $g$ is the acceleration due to gravity, $\omega$ is the frequency, $\left[\phi_{j}(q)\right]=\phi_{j}(q^{+})-\phi_{j}(q^{-})$ is the jump in $\phi$ across the disk, with $q^{+}$ and $q^{-}$ being the ``positive'' and ``negative'' sides of $D$, respectively.
%A explicação detalhada do parâmetro de efeito poroso $G$ foi dada na Introdução (Equação \ref{f4}). 
Furthermore, it is required that $\phi$ satisfies the radiation condition given by \cite{Newman}
\begin{equation}\label{f5}
\lim_{r\rightarrow\infty}{r^{1/2}\left(\frac{\partial \phi_{j}}{\partial r}-iK\phi_{j}\right)}\rightarrow 0,~j=1,\ldots,7,
\end{equation}
where $r=(x^{2}+y^{2})^{1/2}$.

Using Green's theorem, the problem reduces to solving the hypersingular integral equation presented below.
\begin{equation}\label{f6}
\frac{1}{4\pi}{\int{\!\!\!\!\!\!\times}}_{D}{\left[\phi(q)\right]\frac{\partial^{2} {\cal{G}}(p,q)}{\partial n_{p}\partial n_{q}}dA_{q}}-iKG\left[\phi_{j}\right](p)=V_{j},
\end{equation}
where
\begin{equation}\label{f7}
{\cal{G}}(P,Q)\equiv {\cal{G}}(x,y,z;\xi,\eta,\zeta)=(R^{2}+(z-\zeta)^{2})^{-1/2}+{\cal{G}}_{1}(R,z+\zeta),
\end{equation}
\begin{equation}\label{f8}
{\cal{G}}_{1}(R,z+\zeta)={\int{\!\!\!\!\!\!\cup}}_{0}^{\infty}{e^{k(z+\zeta)}J_{0}(kR)\frac{k+K}{k-K}dk},
\end{equation}
in which ${\cal{G}}_{1}$ is a Green function used as the fundamental solution, $R=((x-\xi)^{2}+(y-\eta)^{2})^{1/2}$, $J_{0}$ is a Bessel function, $P=(x,y,z)$ and $Q=(\xi,\eta,\zeta)$, as seen in \cite{farina21}.

\subsection{Reduction to a One-Dimensional Integral Equation}

Following the mathematical framework established in \cite{farinamartin97}, the current objective is to introduce polar coordinates for the field point $p$ and source point $q$, expand the relevant physical quantities into Fourier series regarding the angular variables, and define an unknown auxiliary function $\psi$. This procedure successfully reduces the governing formulation to a one-dimensional integral equation. The systematic sequence of simplifications is detailed as follows.

The field point $p=(x,y)$ is expressed as $p=(r\cos\theta, r\sin\theta)$ within $\Omega$, and similarly, the source point $q=(\xi,\eta)$ is defined as $q=(\rho \cos\varphi, \rho \sin\varphi)$ within $\Omega$, where the area element is $d\Omega = \rho d\rho d\varphi$. By expanding all quantities into Fourier series in the angular variables, equation (\ref{f6}) is transformed into:
\begin{equation}\label{s510}
[f_{n}]_{j}(r)=[V_{n}]_{j}(r)-\frac{1}{2}\int_{0}^{1}{w_{n}(\rho)M_{n}(r,\rho)\rho d\rho}+iKGw_{n}(r),
\end{equation}
where $w_{n}$ denotes the potential jump density satisfying a one-dimensional integral equation. The interaction kernel is defined as $$\displaystyle M_{n}(r,\rho)={\int{\!\!\!\!\!\!\cup}}_{0}^{\infty}{e^{-kb}J_{n}(kr)J_{n}(k\rho)k^{2}\frac{(k+K)}{k-K}dk}.$$ In this context, $[f_{n}]$ represents the radial component of the auxiliary potential acting on the disk, while $[V_{n}]$ represents the radial amplitude of the generalized body velocity for each Fourier mode $n$.

As established by in the literature \cite{Guidera}, $w_{n}$ and $f_{n}$ are related via the following integral identity:
\begin{equation}\label{s511}
w_{n}(r)=-\frac{4}{\pi}r^{n}\int_{r}^{1}{\frac{t^{-2n}}{\sqrt{t^{2}-r^{2}}}}\int_{0}^{t}{\frac{s^{n+1}}{\sqrt{t^{2}-s^{2}}}f_{n}(s)dsdt}, \quad n=0,1,2,\dots
\end{equation}

Substituting (\ref{s510}) into (\ref{s511}) yields:
\begin{equation}\label{s512}
\begin{array}{cc}w_{n}(r)=\displaystyle w_{n}^{\infty}(r)+\int_{0}^{1}w_{n}(\rho)L_{n}(r,\rho)\rho d\rho\\
\displaystyle-\frac{4}{\pi}iKGr^{n}\int_{r}^{1}{\frac{1}{t^{2n}(t^{2}-r^{2})^{1/2}}}\int_{0}^{t}\frac{s^{n+1}}{(t^{2}-s^{2})^{1/2}}w_{n}(s)dsdt,
\end{array}
\end{equation}
where the incident potential term $w_{n}^{\infty}$ and the modified kernel $L_{n}$ are defined as:
\begin{equation}\label{s513}
w_{n}^{\infty}(r)=-\frac{4}{\pi}r^{n}\int_{r}^{1}{\frac{1}{t^{2n}(t^{2}-r^{2})^{1/2}}}\int_{0}^{t}{\frac{s^{n+1}}{(t^{2}-s^{2})^{1/2}}V_{n}(s)dsdt}
\end{equation}
and
\begin{equation}\label{s514}
L_{n}(r,\rho)=\frac{2}{\pi}\rho r^{n}\int_{r}^{1}{\frac{1}{t^{2n}(t^{2}-r^{2})^{1/2}}}\int_{0}^{t}{\frac{s^{n+1}}{(t^{2}-s^{2})^{1/2}}M_{n}(s,\rho)dsdt}.
\end{equation}

A new unknown auxiliary function $\psi_{n}$ is introduced, related to $w_{n}$ through the following integral transformation:
\begin{equation}\label{s61}
w_{n}(r)=D_{n}r^{n}\int_{r}^{1}{\frac{\psi_{n}(t)}{t^{n}(t^{2}-r^{2})^{1/2}}dt},
\end{equation}
where $D_{n}$ is an arbitrary normalization constant. Comparing this relation with (\ref{s511}) yields:
\begin{equation}\label{s62}
\psi_{n}(t)=\psi_{n}^{\infty}(t)+\frac{2}{\pi}t^{-n}\int_{0}^{t}{\frac{s^{n+1}}{(t^{2}-s^{2})^{1/2}}F_{n}(s)ds},
\end{equation}
with the respective terms defined as:
\begin{equation}\label{s63}
\psi_{n}^{\infty}(t)=-\frac{4}{\pi D_{n}}t^{-n}\int_{0}^{t}{\frac{s^{n+1}}{(t^{2}-s^{2})^{1/2}}V_{n}(s)ds}
\end{equation}
and
\begin{equation}\label{s64}
F_{n}(s)=\frac{1}{D_{n}}\left[\left(\int_{0}^{1}{w_{n}(\rho)M_{n}(s,\rho)\rho d\rho}\right)-2iKGw_{n}(s)\right].
\end{equation}

Following the approach established by \cite{farinamartin97}, the governing equation for the potential $\psi_n$ is reformulated as:
\begin{equation}\label{ee1}
\psi_{n}(x)=\psi_{n}^{\infty}(x)-z_{n}(x)-\int_{0}^{1}\psi_{n}(y)N_{n}(x,y)dy,
\end{equation}
where $z_{n}(x)$ represents the porous contribution:
\begin{equation}
    z_{n}(x)=\frac{4}{\pi}iKGx^{-n}\int_{0}^{x}\int_{s}^{1}\frac{s^{2n+1}}{(x^{2}-s^{2})^{1/2}}\frac{\psi_{n}(y)}{y^{n}(y^{2}-s^{2})^{1/2}}dsdy
\end{equation}
and $N_{n}(x,y)$ denotes the free-surface kernel:
\begin{equation}
    N_{n}(x,y)=\frac{2}{\pi}xy{\int{\!\!\!\!\!\!\cup}}_{0}^{\infty}{e^{-kb}j_{n}(kx)j_{n}(ky)k^{2}\frac{k+K}{k-K}dk}.
\end{equation}
This formulation is consistent with the work of \cite{das}, recovering the pure free-surface case when the ice rigidity and inertia parameters vanish.

Under the assumption of purely vertical (heave) oscillations, equation (\ref{ee1}) reduces to the following one-dimensional governing integral equation:
\begin{equation}\label{sss79}
\boxed{
\begin{array}{c}
\displaystyle\psi(x)+\int_{0}^{1}{\psi(y){\cal{K}}(x,y)dy}+\int_{0}^{x}{\psi(y){\cal{I}}(x,y)dy}
+\displaystyle\int_{x}^{1}{\psi(y){\cal{R}}(x,y)dy}=x, \quad 0\leq x\leq 1.
\end{array}}
\end{equation}
The kernels are defined as:
\begin{equation}
{\cal{K}}(x,y)=-N_{0}(x,y)+\frac{4}{\pi}iKG\ln(|x|+|y|),
\end{equation}
\begin{equation}
{\cal{I}}(x,y)=-\frac{2}{\pi}iKG\ln(x^{2}-y^{2}),
\end{equation}
\begin{equation}
{\cal{R}}(x,y)=-\frac{2}{\pi}iKG\ln(y^{2}-x^{2}),
\end{equation}
and
\begin{equation}
N_{0}(x,y)=\frac{b}{\pi}(b^{2}+X^{2})^{-1}+\frac{2K}{\pi}\Phi_{0}(X,b)-\frac{b}{\pi}(b^{2}+T^{2})^{-1}-\frac{2K}{\pi}\Phi_{0}(T,b),
\end{equation}
where $X=x-y$ and $T=x+y$. Here, $\Phi_{0}$ is a two-dimensional wave potential that can be efficiently evaluated using the expansion provided by \cite{yu61}:
\begin{equation}\label{s776}
\begin{aligned}
\Phi_{0}(B,C) = & -e^{-KC} \left[ (\ln(KS) - i\pi + \gamma)\cos(KB) + \beta \sin(KB) \right] \\
& + \sum_{m=1}^{\infty} \frac{(-KS)^{m}}{m!} \left( \sum_{k=1}^{m} \frac{1}{k} \right) \cos(m\beta).
\end{aligned}
\end{equation}
This model preserves the functional dependence on the wavenumber $K$, the porosity parameter $G$, and the submergence $d=b/2$.

\subsection{The Connection with the Love-Lieb Equation}
In this section, the simplifications that emerge from the problem formulation under specific conditions, notably when the wave number $K$ is zero, are explored. This particular case reveals a direct connection with the classic Love equation, a fundamental problem in potential theory that describes the behavior of submerged discs in an incompressible and irrotational fluid. By examining this reduction, the consistency of the approach under analysis is not only validated, but a starting point for future investigations into the generalizations of the Love equation, particularly relevant for fluid-structure interaction, is also established.

Indeed, when $K=0$, the governing integral equation (\ref{sss79}) simplifies to
\begin{equation}\label{eeqn2}
\psi(x)+\int_{0}^{1}{\psi(y)\left[-N_{0}(x,y)\right]dy}=x,~0\leq x\leq 1.
\end{equation}

Considering the expression for $N_{0}(x,y)$ given in (\ref{sss79}), this equation can be explicitly rewritten as
\begin{equation}\label{eqn11}
\psi(x)-\frac{b}{\pi}\int_{0}^{1}\frac{\psi(y)}{b^{2}+(x-y)^{2}}dy+\frac{b}{\pi}\int_{0}^{1}\frac{\psi(y)}{b^{2}+(x+y)^{2}}dy=x,~0\leq x\leq 1,
\end{equation}
where $d=\frac{b}{2}$.

Equation (\ref{eqn11}) shows a notable similarity with the family of integral equations known as the Love-Lieb equations. More precisely, it resembles the simplest form, ($L_{1}^{\pm}$), presented in \cite{love22}, especially when restricted to the interval $0\leq x\leq 1$. The imposition of $K=0$ on the free surface condition of the problem under analysis is physically equivalent to the presence of a mirrored disc, which, in terms of an electrostatic problem, would correspond to the configuration of two coaxial discs separated by a distance $d$ \citep{love22}.

It is important to emphasize that, to date, no closed-form analytical solution is known for the Love-Lieb equations, which consolidates them as a fundamental and challenging problem in potential theory. Equation (\ref{eqn11}), therefore, represents a natural generalization of this classical formulation. This generalization not only allows for the adaptation of the problem to different hydrodynamic contexts but can also be interpreted as an alternative form of the Love-Lieb equations, expanding their scope of application in problems involving fluid-structure interaction.

\section{NUMERICAL METHOD}

Equation (\ref{sss79}) is formulated as a Fredholm integral equation of the second kind featuring a singular kernel. The numerical treatment of these inherent singularities was addressed using the Boundary Element Method (BEM) following \cite{kaitai} coupled with the \texttt{D01GCF} routine from the \text{NAG} library \citep{nag1}, which is based on the rigorous evaluation of potential integrals \citep{farina2001evaluation}. This routine employs the Korobov-Conroy Number-Theoretic Method (NTM) to approximate definite integrals in up to 20 dimensions; further theoretical details regarding this method are available in \cite{kaitai}.

The strategy to mitigate the singularity involved decomposing the original kernel into two distinct components, followed by partitioning the integration domain into subintervals adjacent to the singular point. Under this framework, the resulting integrals were evaluated using the \texttt{D01GCF} routine. The discretization procedure partitioned the interval $[0,1]$ into $n$ finite elements, denoted as $I_{j}=[(j-1)/n, j/n]$, adopting the midpoint as the collocation node, defined by $x_j = (2j-1)/(2n)$.

Similarly, the intervals $\left[0,x\right]$ and $\left[x,1\right]$ are discretized using the same subintervals, with the index $p$ running from $1$ to $j$ and the index $q$ running from $j+1$ to $n$.
Note that $I_{p}\cup I_{q}=\left[0,1\right]$. 
The collection of all subintervals indexed by $p$ and $q$ covers the entire domain $[0,1]$.
With this in hand, (\ref{sss79}) is rewritten as
\begin{equation}\label{tsn211}
\begin{array}{cc}\displaystyle\psi(x)+\sum_{j=1}^{n}{\int_{I_{j}}{\psi(y){\cal{K}}(x,y)dy}}+\sum_{p=1}^{j}\int_{I_{p}}{\psi(y){\cal{I}}(x,y)dy}
\displaystyle+\sum_{q=j+1}^{n}\int_{I_{q}}{\psi(y){\cal{R}}(x,y)dy}=x,
\end{array}
\end{equation}
and evaluated at each collocation point $x_{i}$, $i=1,2,3,\ldots,n$.
\begin{equation}\label{ttsn211}
\psi(x_{i})+\sum_{j=1}^{n}{\int_{I_{j}}{\psi(y){\cal{K}}(x_{i},y)dy}}+\sum_{p=1}^{j}{\int_{I_{p}}{\psi(y){\cal{I}}(x_{i},y)dy}}
+\sum_{q=j+1}^{n}{\int_{I_{q}}{\psi(y){\cal{R}}(x_{i},y)dy}}=x_{i}
\end{equation}
It is now assumed that $\psi$ is a step function of the type
\begin{equation}\label{s714}
\psi(x)=\left\{
\begin{array}{cc}\psi_{i},~x\in\left[\frac{i-1}{n},\frac{i}{n}\right)\\0,~x\notin\left[\frac{i-1}{n},\frac{i}{n}\right)
\end{array}.\right.
\end{equation}
Thus, (\ref{tsn211}) can be approximated by
\begin{equation}\label{s715}
\psi(x_{i})+\sum_{j=1}^{n}{\psi_{j}\int_{I_{j}}{{\cal{K}}(x_{i},y)dy}}+\sum_{p=1}^{j}{\psi_{p}\int_{I_{p}}{{\cal{I}}(x_{i},y)dy}}
+\sum_{q=j+1}^{n}{\psi_{q}\int_{I_{q}}{{\cal{R}}(x_{i},y)dy}}=x_{i}
\end{equation}
Writing equation (\ref{s715}) in matrix form, we obtain
\begin{equation}\label{s716}
(I+A+B+C)\psi=x,
\end{equation}
where $I$ is the $n \times n$ identity matrix, $\psi = [\psi_1 \psi_2 \ldots \psi_n]^T$ is the vector of unknown coefficients, $x = [x_1 x_2 \ldots x_n]^T$ is the vector of collocation points, $A=\left[a_{i,j}\right]_{i,j=1}^{n}$ is a matrix, $B=\left[b_{i,j}\right]_{i,j=1}^{n}$ is a lower triangular matrix, $C=\left[c_{i,j}\right]_{i,j=1}^{n}$ is an upper triangular matrix, and
$$A=\int_{I_{j}}{\left[-N_{0}(x,y)+\frac{4}{\pi}iKG\ln(|x|+|y|)\right]dy},
$$
$$
B=\int_{I_{p}}{-\frac{2}{\pi}iKG\ln(x^{2}-y^{2})dy},~C=\int_{I_{q}}{-\frac{2}{\pi}iKG\ln(y^{2}-x^{2})dy}
$$
$$
N_{0}(x_{i},y)=\frac{b}{\pi}(b^{2}+(x_{i}-y)^{2})^{-1}+\frac{2K}{\pi}\Phi_{0}(x_{i}-y,b)-\frac{b}{\pi}(b^{2}+(x_{i}+y)^{2})^{-1}-\frac{2K}{\pi}\Phi_{0}(x_{i}+y,b).
$$
In this way, we obtain a linear system that we can solved numerically. With the obtained value of $\psi$, the added mass and damping coefficient are calculated as indicated in \cite{falnes2002},
\begin{equation}\label{s7771}
{\cal A}(K,b)+i{\cal B}(K,b)=8\int_{0}^{1}{\psi(x)x}dx,
\end{equation}
where ${\cal A}$ is the added mass and ${\cal B}$ is the damping coefficient and $d=\frac{b}{2}$.

\section{RESULTS AND DISCUSSION}
The linear system derived from the discretization of the governing equation was implemented in FORTRAN, with simulations conducted following the methodology proposed by \cite{farina21} for benchmarking purposes. Model validation included a grid independence test, resulting in the selection of a quadrilateral mesh with $N = 80$. The choice of this discretization level was based on convergence analysis and is supported by the methodological precedent of \cite{das22}, who employed an identical parameter in a related study. This configuration proved adequate to ensure the numerical precision required for the subsequent analyses.
\begin{table}[htb!]
\centering
\caption{Mesh independence - $d=0.1$ and $K=0.5$.}
\label{exx2}
\begin{tabular}{ccccccc}
\hline
N & $G$ & ${\cal{A}}$ & ${\cal{B}}$& $G$ & ${\cal{A}}$ & ${\cal{B}}$ \\
\hline
50 & $G=0.1$ & $-8.6628$ & $9.8061$& $G=0.1i$ & $-9.2606$ & $9.3202$ \\
60 & $G=0.1$ & $-8.6640$ & $9.8263$& $G=0.1i$ & $-9.2543$ & $9.3356$ \\
70 & $G=0.1$ & $-8.6715$ & $9.8440$& $G=0.1i$ & $-9.2495$ & $9.3480$ \\
80 & $G=0.1$ & $-8.6720$ & $9.8558$& $G=0.1i$ & $-9.2471$ & $9.3530$ \\
90 & $G=0.1$ & $-8.6720$ & $9.8558$& $G=0.1i$ & $-9.2471$ & $9.3530$ \\
\hline
\end{tabular}
\end{table}
\begin{table}[htb!]
\centering
\caption{Mesh independence - $d=0.2$ and $K=0.5$.}
\label{exx3}
\begin{tabular}{cccccccccc}
\hline
N & $G$ & ${\cal{A}}$ & ${\cal{B}}$& $G$ & ${\cal{A}}$ & ${\cal{B}}$ & $G$ & ${\cal{A}}$ & ${\cal{B}}$ \\
\hline
50 & $G=0$ & $4.7745$ & $8.6131$& $G=1$ & $3.2636$ & $5.0759$ & $G=0.7+0.3i$ & $3.7538$ & $6.8198$\\
60 & $G=0$ & $4.7669$ & $8.6120$& $G=1$ & $3.2751$ & $5.0647$ & $G=0.7+0.3i$ & $3.7622$ & $6.8062$\\
70 & $G=0$ & $4.7631$ & $8.6118$& $G=1$ & $3.2812$ & $5.0628$ & $G=0.7+0.3i$ & $3.7752$ & $6.8066$\\
80 & $G=0$ & $4.7622$ & $8.6111$& $G=1$ & $3.2831$ & $5.0617$ &$G=0.7+0.3i$ & $3.7880$ & $6.8097$ \\
90 & $G=0$ & $4.7622$ & $8.6111$& $G=1$ & $3.2831$ & $5.0617$ &$G=0.7+0.3i$ & $3.7880$ & $6.8097$\\
\hline
\end{tabular}
\end{table}

The results of the problem under study are presented next, detailing the Added Mass (${\cal{A}}$) and Damping Coefficient (${\cal{B}}$). These coefficients were calculated for a variety of $G$ values, which will be presented later. For the analysis, the five scenarios proposed in \cite{farina21} were considered, with the respective visual representations included for each case (Figures 4 to 8). However, to illustrate the consistency and robustness of the proposed formulation, Figures 2 and 3 present a direct comparison between the Added Mass and Damping Coefficients obtained in the current study and the corresponding results presented by \cite{farina21}. This initial visual analysis highlights the close proximity between the curves, providing a preliminary validation of the approach under analysis before the detailed discussion of the specific scenarios.
\begin{figure}[htb!]
\hspace*{-1cm}
\centering
\includegraphics[width=1\textwidth]{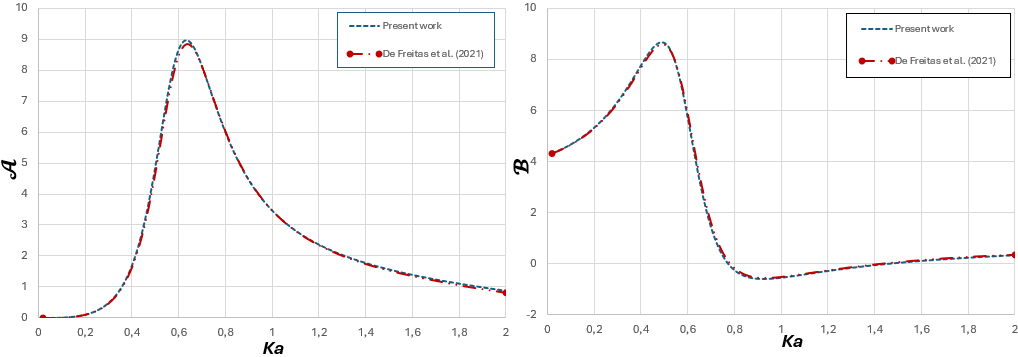}
\caption{ {\small Added mass and damping coefficients with $G=0$ and $d=0.2$.}}
\label{figura2}
\end{figure}
\begin{figure}[htb!]
\hspace*{-1cm}
\centering
\includegraphics[width=1\textwidth]{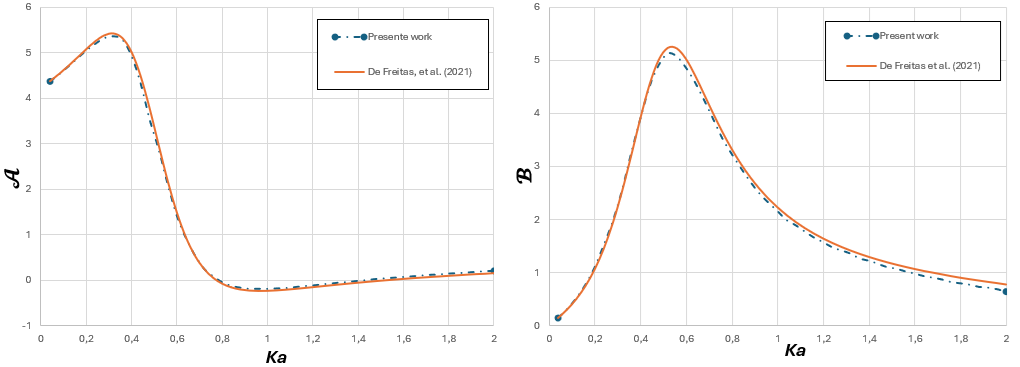}
\caption{ {\small Added mass and damping coefficients with $G=1$ and $d=0.2$.}}
\label{figura3}
\end{figure}

\subsection{Case 1: Variable Imaginary $G$ and Fixed $d$}
\begin{figure}[htb!]
\hspace*{-1cm}
\centering
\includegraphics[width=1\textwidth]{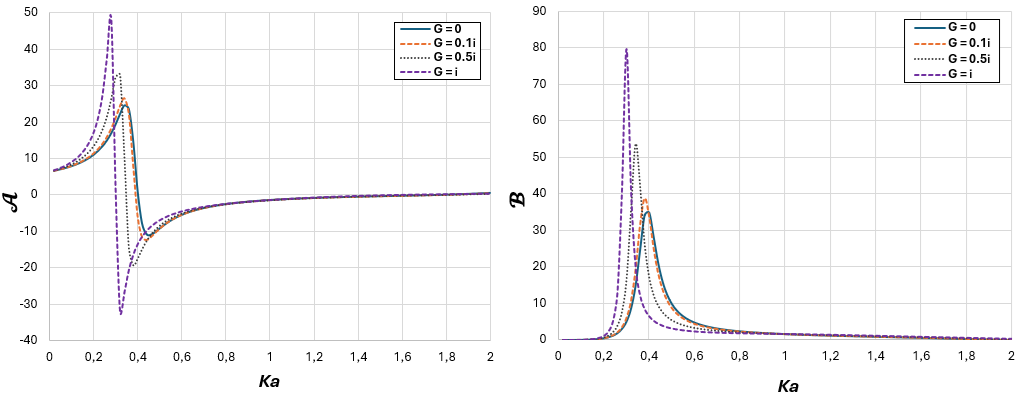}
\caption{ {\small Added mass and damping coefficients as a function of $Ka$, for $d=0.1$ and imaginary values of $G$.}}
\label{figura4}
\end{figure}

Considering $G$ as a purely imaginary number taking 4 distinct values and the fixed depth, $d=0.1$, simulations were performed and the graphs obtained, as shown in Figure \ref{figura4}.\\\\\\\\\\\\\\\\

The results obtained through the formulation under analysis demonstrate a notable agreement with those presented by \cite{farina21}, in Figures 8 and 10. This graphical similarity, observed for the specific case where $G$ is purely imaginary, suggests the validity of the proposed formulation, indicating a consistent behavior with the results previously established in the literature.
\subsection{Case 2: Variable Real $G$ and Fixed $d$}
\begin{figure}[htb!]
\hspace*{-1cm}
\centering
\includegraphics[width=1\textwidth]{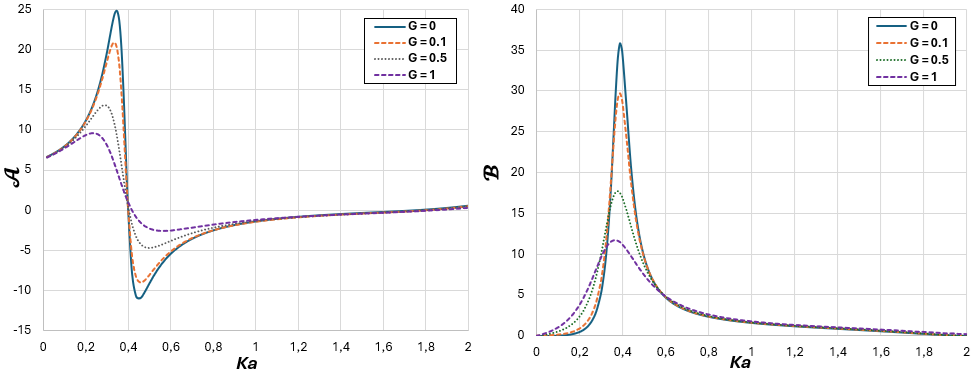}
\caption{ {\small Added mass and damping coefficients as a function of $Ka$, for $d=0.1$ and real values of $G$.}}
\label{figura5}
\end{figure}

Considering $G$ as a real number taking 4 distinct values and the fixed depth, $d=0.1$, simulations were performed and the graphs obtained, as shown in Figure \ref{figura5}.

It can be noted that the graphs presented by the formulation under analysis closely approach what was presented in \cite{farina21}, Figures 9 and 11, showing that the formulation found, for the case where $G$ is real, presents a compatible result.
\subsection{Case 3: $G=0$ and Variable $d$}
\begin{figure}[htb!]
\hspace*{-1cm}
\centering
\includegraphics[width=1\textwidth]{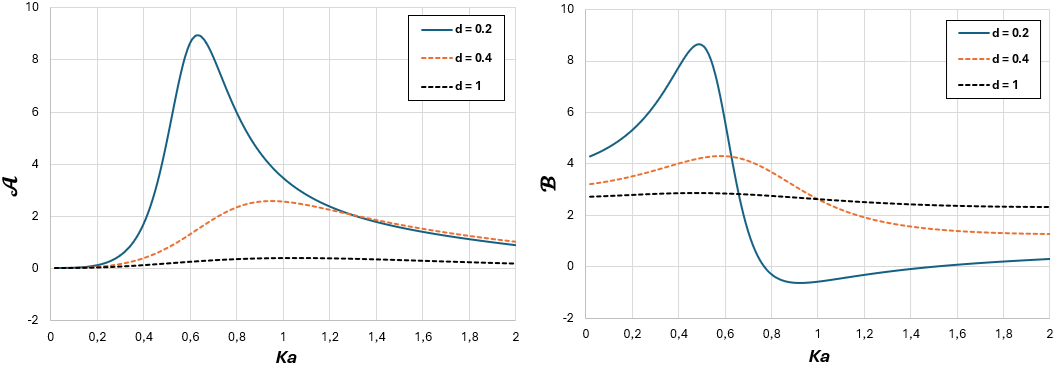}
\caption{ {\small Added mass and damping coefficients as a function of $Ka$, for different values of $d$ and $G=0$.}}
\label{figura6}
\end{figure}

Considering $G=0$ and the depth $d$ taking three distinct values, simulations were performed and the graphs obtained, as shown in Figure \ref{figura6}.

It can be noted that the graphs presented by the formulation under analysis closely approach what was presented in \cite{farina21}, Figures 2 and 3, showing that the formulation found, for the case where $G=0$, presents a compatible result.

\subsection{Case 4: $G=1$ and Variable $d$}
\begin{figure}[htb!]
\hspace*{-1cm}
\centering
\includegraphics[width=0.93\textwidth]{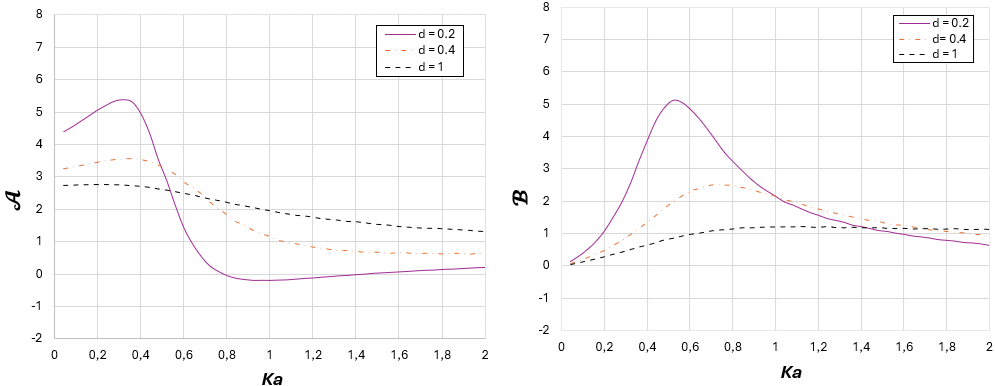}
\caption{{\small Added mass and damping coefficients as a function of $Ka$, for different values of $d$ and $G=1$.}}
\label{figura7}
\end{figure}

Considering $G=1$ and the depth $d$ taking four distinct values, simulations were performed and the graphs obtained, as shown in Figure \ref{figura7}.\\\\\\\\\\\\\\

It can be noted that the graphs presented by the formulation under analysis closely approach what was presented in \cite{farina21}, Figures 6 and 7, validating the research under analysis for the case where $G=1$.

\subsection{Case 5: $G=0.7+0.3i$ and Variable $d$}
Considering $G=0.7+0.3i$ and the depth $d$ taking four distinct values, simulations were performed and the graphs obtained, as shown in the image below.
\begin{figure}[htb!]
\hspace*{-1cm}
\centering
\includegraphics[width=1\textwidth]{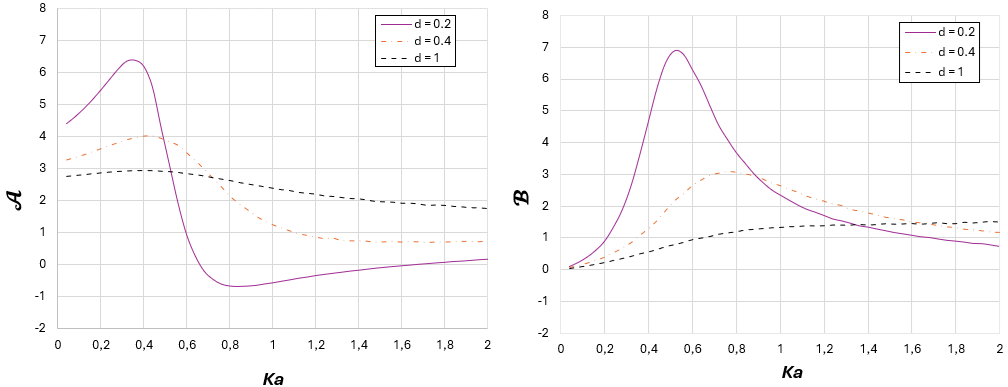}
\caption{{\small Added mass and damping coefficients as a function of $Ka$, for different values of $d$ and $G=0.7+0.3i$.}}
\label{figura8}
\end{figure}

It can be noted that the graphs presented by the formulation under analysis closely approach what was presented in \cite{farina21}, Figures 4 and 5, validating the research under analysis for the case where $G=0.7+0.3i$.

To reinforce the good agreement between the results of this research and those of \cite{farina21}, the Mean Absolute Error (MAE) and the Root Mean Square Error (RMSE) were used to quantify the differences between the values of the studied problem and the values from \cite{farina21}, taking one graph (values) for each of the presented cases, as shown in Tables 3 and 4 and based on \cite{dados}.
\begin{table}[h!]
\centering
\caption{Quantifying graphical results via RMSE and MAE.}
\label{exxxx}
\begin{tabular}{ccccc}
\hline
~&\multicolumn{2}{c}{$G=0.1$ and $d=0.1$}&\multicolumn{2}{c}{$G=0.1i$ and $d=0.1$}\\\hline
Metric&${\cal{A}}$&${\cal{B}}$&${\cal{A}}$&${\cal{B}}$\\\hline
RMSE&0.137&0.141&0.30&0.24\\
MAE&0.101&0.063&0.17&0.11\\\hline
\end{tabular}
\end{table}
\begin{table}[h!]
\centering
\caption{Quantifying graphical results via RMSE and MAE.}
\label{exx4}
\begin{tabular}{ccccccc}
\hline
~&\multicolumn{2}{c}{$G=0$ and $d=0.2$}&\multicolumn{2}{c}{$G=1$ and $d=0.2$}&\multicolumn{2}{c}{$G=0.7+0.3i$ and $d=0.2$}\\\hline Metric&${\cal{A}}$&${\cal{B}}$&${\cal{A}}$&${\cal{B}}$&${\cal{A}}$&${\cal{B}}$\\\hline
RMSE&0.0926&0.0951&0.0741&0.0797&0.0849&0.0658\\
MAE&0.056&0.070&0.0511&0.0692&0.047&0.055\\\hline
\end{tabular}
\end{table}

\section{CONCLUSION}
The results obtained throughout this research demonstrate that the sequence of simplifications proposed in \cite{farinamartin97}, originally aimed at deriving a one-dimensional equation for wave interaction with a rigid circular plate, was successfully adapted to the context of porous bodies. The same analytical strategy proved effective in obtaining a streamlined one-dimensional formulation for a porous circular disk while preserving the essential mathematical structure of the original model.

As discussed in Section 4 and evidenced by Figures 2 to 8, the numerical results show strong agreement with those reported in \cite{farina21}. Such compatibility reinforces the validity of the proposed integral formulation, particularly regarding added mass and damping coefficients. For future work, it is suggested to refine the approximation techniques to circumvent numerical limitations encountered during this study, noting that the current investigation focused on small submergence depths, thus excluding the $d=10$ cases.

The analysis allowed for a deeper investigation into the influence of porosity on hydrodynamic interaction by varying the complex porous parameter $G$. When $G$ assumes purely real values, it physically represents a high-resistance porous structure. Under this condition, as real $G$ increases, equivalent to a decrease in resistance $f$ in equation (\ref{f4}), the disk behaves increasingly like a solid plate, hindering fluid penetration and consequently reducing added mass and damping effects, as shown in Figure 5.

Conversely, when $G$ is purely imaginary, the system is dominated by the inertial effect represented by the parameter $S$ (see equation (\ref{f4})). This implies that the fluid within the pores tends to follow the disk's motion, acting as an effective added mass coupled to the structure. This effect is particularly relevant in near-free-surface configurations, which can induce resonance in the fluid layer above the disk. The presence of inertial-dominant porosity ($G$ being purely imaginary) can amplify this resonance, leading to significantly high peaks in both added mass and damping coefficients, as presented in Figure 4.

From a physical standpoint, the system's response as a function of $G$ reflects the balance between the resistive and inertial components of the porous flow. High values of the real part of $G$ indicate a more resistant medium, leading to flow blockage and concentrated hydrodynamic forces on the structure. On the other hand, high values of the imaginary part of $G$ reflect a predominantly inertial response, favoring relative flow and reducing viscous resistance.

Furthermore, it was observed that for $K=0$, equation (\ref{eqn11}) resembles the classical Love equation, originally associated with the electrostatic problem of a capacitor with coaxial circular plates. This formulation allows for a generalization to hydrodynamic contexts and can be interpreted as a variation of the Love-Lieb equations \citep{love22}, contributing to fluid-structure interaction studies across various applied mathematical modeling scenarios.

A significant limitation concerning the numerical implementation of the governing equation was identified. In contrast to the approach in \cite{farinamartin97}, the direct application of the Nyström method with Gauss-Legendre quadrature proved insufficient. The singularity inherent to equation (\ref{sss79}) hindered the use of standard methodologies, necessitating the development of an alternative numerical strategy for integral evaluation. This difficulty in treating singularities represents a path for future research in the numerical analysis of singular integral equations.

In conclusion, the research objectives were fully met. It was possible to derive an integral formulation for porous plates following the methodology in \cite{farinamartin98}. The numerical consistency validates the developed approach and provides a better understanding of the physical effects introduced by porosity on the system's hydrodynamic response.

%\begin{quote}
%\textbf{Mathematical Phenomenon Highlight: Resonance in the Small Submergence Regime} \\
In the small submergence regime ($d = 0.1$), the analysis focuses on resonance phenomena intensified by the proximity of the free surface. The nature of the porosity parameter $G$ significantly alters the hydrodynamic response: while an increase in the imaginary component ($G_i$) raises added mass peaks and shifts resonance frequencies to lower values, an increase in the real component ($G_r$) acts dissipatively, reducing peak magnitudes. Mathematically, this regime is governed by the oscillation of the thin fluid layer above the disk, where complex porosity serves as a tuning mechanism for controlling the extreme values of incident hydrodynamic forces.
%\end{quote}

%----------------------------------------------------------------------------------------
%	REFERENCES
%----------------------------------------------------------------------------------------
\interlinepenalty=10000
\renewcommand{\refname}{REFERENCES}
\bibliographystyle{apalike}
\sloppy\bibliography{references}

@article{farina2001evaluation,
  title={Evaluation of single layer potentials over curved surfaces},
  author={Farina, Leandro},
  journal={SIAM Journal on Scientific Computing},
  volume={23},
  number={1},
  pages={81--91},
  year={2001},
  publisher={SIAM}
}

@article{martin1997interaction,
  title={Interaction of water waves with thin plates},
  author={Martin, PA and Parsons, NF and Farina, L},
  journal={International Series on Advances in Fluid Mechanics},
  volume={8},
  pages={197--230},
  year={1997},
  publisher={Computational Mechanics Publications}
}

@article{chwangat94,
  AUTHOR ={Chwang, A. T. and Wu, J.},
  TITLE ={Wave scattering by submerged porous disk},
  JOURNAL=    {Journal of Engineering Mechanics},
  volume={120},
  YEAR ={1994},
}

@article{farinamartin98,
  title={Scattering of Water Waves by a Submerged Disc Using a Hipersingular Integral Equation},
  author={Farina, Leandro and Martin, P. A.},
  journal={Applied Ocean Research},
  volume={20},
  pages={121--134},
  year={1998},
}

@article{antonutti14,
  title={An investigation of the effects of wind-induced inclination on floating wind turbine dynamics: heave plate excursion},
  author={Antonutti, R. and Peyrard, C. and Johanning, L. and Incecik, A. and Ingram, D.},
  journal={Ocean Engineering},
  volume={91},
  pages={208--217},
  year={2014},
}

@article{astariz15,
  title={The economics of wave energy: A review},
  author={Astariz, S. and Iglesias, G.},
  journal={Renewable and Sustainable Energy Reviews},
  volume={45},
  pages={397--408},
  year={2015},
}

@article{chwang1983,
  title={A porous-wavemaker theory},
  author={Chwang, A. T.},
  journal={Journal of Fluid Mechanics},
  volume={132},
  pages={395--406},
  year={1983},
}

@article{farina21,
  title={The heaving motion of a porous disc submerged in deep water},
  author={De Freitas, I. M. and Farina, L and Miller, J. J. H.},
  journal={Ocean Engineering},
  volume={219},
  pages={108--290},
  year={2021},
}

@article{farinamartin97,
  title={Radiation of water waves by a heaving submerged horizontal disc},
  author={Farina, Leandro and Martin, P. A.},
  journal={Journal of Fluid Mechanics},
  volume={337},
  pages={365--379},
  year={1997},
}

@article{dokken17,
  title={Wave analysis of porous geometry with linear resistance law},
  author={Dokken, J. and Grue, J. and Karstensen, L.P.},
  journal={Journal of Marine Science and Application},
  volume={16},
  pages={480--489},
  year={2017},
}

@article{gayen14,
  title={A hypersingular integral equation approach to the porous plate problem},
  author={Gayen, R. and Mondal, A.},
  journal={Applied Ocean Research},
  volume={46},
  pages={70--78},
  year={2014},
}

@article{gayen16,
  title={Water wave interaction with two symmetric inclined permeable plates},
  author={Gayen, R. and Mondal, A.},
  journal={Ocean Engineering},
  volume={124},
  pages={180--191},
  year={2016},
}

@article{Lamas,
  title={A review of very large floating structures (VLFS) for coastal and offshore uses},
  author={Lamas-Pardo, M. and Iglesias, G. and Carral, L.},
  journal={Ocean Engineering},
  volume={109},
  pages={677--690},
  year={2015},
}

@article{LopezSouto,
  title={Hydrodynamic coefficients and pressure loads on heave plates for semi-submersible floating offshore wind turbines: a comparative analysis using large scale models},
  author={Lopez-Pavon, C. and Souto-Iglesias, A.},
  journal={Renewable Energy},
  volume={81},
  pages={864--881},
  year={2015},
}

@article{Sollit,
  title={Wave transmission through permeable breakwaters},
  author={Sollitt, C.K. and Cross, R.H.},
  journal={Coastal Engineering Proceedings},
  volume={1(13)},
  pages={p.99},
  year={1972},
}

@article{Koley,
  title={Fredholm integral equation technique for hydroelastic analysis of a floating flexible porous plate},
  author={Koley, S. and Mondal, R. and Sahoo, T.},
  journal={European Journal of Mechanics/B Fluids},
  volume={67},
  pages={291--305},
  year={2018},
}

@article{Liu11,
  title={A new approximate analytic solution for water wave scattering by a submerged horizontal porous disk},
  author={Yong Liu and Hua-jun Li and Yu-cheng Li and Shi-yan He},
  journal={Applied Ocean Research},
  volume={67},
  pages={286--296},
  year={2011},
}

@article{Molin,
  title={ Heave added mass and damping of a perforated disk below the free surface},
  author={Molin, B. and Nielsen, F.G.},
  journal={Proceedings of the 19th International Workshop on Water Waves and Floating Bodies},
  year={2004}
}

@article{Ouled,
  title={ On wave diffraction-radiation by bodies with porous thin plates},
  author={Ouled Housseine, C.},
  journal={Presented at 34th International Workshop on Water Waves and Floating Bodies},
  year={2019}
}

@article{Mackay,
  title={Verification of a Boundary Element Model for Wave Forces on Structures with Porous Elements},
  author={Mackay, E.B.L. and Feichtner, A. and Smith, R.E. and Thies, P.R. and Johanning, L.},
  journal={Advances in Renewable Energies Offshore - Proceedings of the 3rd International Conference on Renewable Energies Offshore},
  pages={341--350},  
year={2019},
}

@book{Newman,
  author = {Newman, J. N.},
  title = {Marine Hydrodynamics}, 
  series = {Springer Briefs in Mathematics},
  publisher = {The MIT Press},
  year = {1977},
  isbn = {9780262280617},
 doi = {10.7551/mitpress/4443.001.0001}
}

@article{FARINA2017165,
title = {Radiation of water waves by a submerged nearly circular plate},
journal = {Journal of Computational and Applied Mathematics},
volume = {310},
pages = {165--173},
year = {2017},
note = {Numerical Algorithms for Scientific and Engineering Applications},
issn = {0377-0427},
doi = {https://doi.org/10.1016/j.cam.2016.04.009},
url = {https://www.sciencedirect.com/science/article/pii/S0377042716301819},
author = {Leandro Farina and Rômulo L. {da Gama} and Sergey Korotov and Juliana S. Ziebell},
}

@article{garrett1971,
title={Wave forces on a circular dock},
volume={46},
DOI={10.1017/S0022112071000430},
number={1},
journal={Journal of Fluid Mechanics},
publisher={Cambridge University Press},
author={Garrett, C. J. R.}, year={1971},
pages={12--139}}

@article{Guidera,
  title={Penny-shaped cracks},
  author={Guidera, J. T. Lardner, R. W.},
  journal={Journal of Elasticity},
  volume={5},
  pages={59--73},
  year={1975},
 doi = {10.1007/BF01389258}
}

@article{yu61, 
title={Surface waves generated by an oscillating circular cylinder on water of finite depth: theory and experiment}, 
volume={11}, 
DOI={10.1017/S0022112061000718}, 
number={4}, 
journal={Journal of Fluid Mechanics}, 
publisher={Cambridge University Press}, 
author={Yu, Y. S. and Ursell, F.}, 
year={1961}, 
pages={529--551}}

@book{falnes2002,
place={Cambridge},
title={Ocean Waves and Oscillating Systems: Linear Interactions Including Wave-Energy Extraction},
DOI={10.1017/CBO9780511754630},
publisher={Cambridge University Press},
author={Falnes, Johannes},
year={2002}}

@article{xiping94b, 
title={Wave-induced oscillation in harbor with porous breakwaters}, 
volume={120}, 
number={2}, 
journal={Journal of Waterway, Port, Coastal, and Ocean Engineering}, 
author={Yu, X. and Chwang, A.T.}, 
year={1994}, 
pages={125--144}}

@article{xiping95, 
title={Diffraction of water waves by porous breakwaters}, 
volume={121}, 
number={6}, 
journal={Journal of Waterway, Port, Coastal, and Ocean Engineering}, 
author={Yu, X.}, 
year={1995}, 
pages={275--282}}

@article{Parsons1995, 
title={Trapping of water waves by submerged plates using hypersingular integral equations}, 
journal={Journal of Fluid Mechanics},
volume={284}, 
pages={359–375},
year={{1995}},
DOI={10.1017/S0022112095000395},  
author={Parsons, Neil F. and Martin, P. A.}, 
}

@article{kaitai,
 ISSN = {08834237},
 URL = {http://www.jstor.org/stable/2246355},
 author = {Kai-Tai Fang and Yuan Wang and Peter M. Bentler},
 journal = {Statistical Science},
 number = {3},
 pages = {416--428},
 publisher = {Institute of Mathematical Statistics},
 title = {Some Applications of Number-Theoretic Methods in Statistics},
 urldate = {2023-11-13},
 volume = {9},
 year = {1994}
}

@article{das,
title = {Radiation of water waves by a heaving submerged disc in a three-layer fluid},
journal = {Journal of Fluids and Structures},
volume = {111},
pages = {103575},
year = {2022},
issn = {0889-9746},
doi = {https://doi.org/10.1016/j.jfluidstructs.2022.103575},
url = {https://www.sciencedirect.com/science/article/pii/S0889974622000469},
author = {Arijit Das and Soumen De and B.N. Mandal},
}

@article{islam,
author = {Islam, Najnin and Kundu, Souvik and Gayen, Rupanwita},
year = {2019},
month = {12},
pages = {20190331},
title = {Scattering and radiation of water waves by a submerged rigid disc in a two-layer fluid},
volume = {475},
journal = {Proceedings of The Royal Society A Mathematical Physical and Engineering Sciences},
doi = {10.1098/rspa.2019.0331}
}

@article{love22,
  title={Love--{L}ieb Integral Equations: Applications, Theory, Approximations, and Computations},
  author={Farina, Leandro and Lang, Guillaume and Martin, PA},
  journal={SIAM Review},
  volume={64},
  number={4},
  pages={831--865},
  year={2022},
  publisher={SIAM}
}

@article{PARSONS1994,
title = {Scattering of water waves by submerged curved plates and by surface-piercing flat plates},
journal = {Applied Ocean Research},
volume = {16},
number = {3},
pages = {129-139},
year = {{1994}},
issn = {0141-1187},
doi = {https://doi.org/10.1016/0141-1187(94)90024-8},
url = {https://www.sciencedirect.com/science/article/pii/0141118794900248},
author = {N.F. Parsons and P.A. Martin},
}

@ARTICLE{YuC94,
	author = {Yu, Xiping and Chwang, Allen T.},
	title = {Water waves above submerged porous plate},
	year = {1994},
	journal = {Journal of Engineering Mechanics},
	volume = {120},
	number = {6},
	pages = {1270 – 1282},
	doi = {10.1061/(ASCE)0733-9399(1994)120:6(1270)},
	type = {Article},
	publication_stage = {Final},
	source = {Scopus},
	note = {Cited by: 116}
}

@article{ZIEE,
title = {Water wave radiation by a submerged rough disc},
journal = {Wave Motion},
volume = {49},
number = {1},
pages = {34-49},
year = {2012},
issn = {0165-2125},
doi = {https://doi.org/10.1016/j.wavemoti.2011.07.001},
url = {https://www.sciencedirect.com/science/article/pii/S0165212511000825},
author = {Juliana S. Ziebell and Leandro Farina}}

@article{dados,
author  = 
{Carmo, Carlos Roberto Souza  and Silva, Jéssica Rayse de Melo},
 title = {APRENDIZADO DE MáQUINA E PRESTAçãO DE SERVIçOS DE
ARMAZENAMENTO DE DADOS: Métricas PARA Análise E
VALIDAçãO DE ALGORITMOS PREVISORES},
journal = {GETEC},
volume  = {12},
year = {2023},
page = {123--144},
}

@article{das22,
author = {Das, Arijit and De, Soumen and Mandal, B.},
year = {2022},
month = {04},
pages = {1557-1573},
title = {Radiation and scattering of flexural-gravity waves by a submerged porous disc},
volume = {57},
journal = {Meccanica},
doi = {10.1007/s11012-022-01510-y}
}

@article{PARSONS1992,
title = {Scattering of water waves by submerged plates using hypersingular integral equations},
journal = {Applied Ocean Research},
volume = {14},
number = {5},
pages = {313-321},
year = {{1992}},
issn = {0141-1187},
doi = {https://doi.org/10.1016/0141-1187(92)90035-I},
url = {https://www.sciencedirect.com/science/article/pii/014111879290035I},
author = {N.F. Parsons and P.A. Martin},
}

@manual{nag1,
  title        = {D01GCF: One-Dimensional Quadrature (General-Purpose Integrator)},
  author       = {{The Numerical Algorithms Group}},
  year         = {n.d},
  note         = {NAG Library Manual},
url ={https://support.nag.com/numeric/fl/nagdoc_latest/html/d01/d01gcf.html},
organization = {The Numerical Algorithms Group}
}

@book{farina12,
  title={Ondas Oceânicas de Superfície},
  author={Farina, L.},
   edition =      {2th},
  year={2012},
  publisher={SBMAC}
}

\end{document}